\documentstyle[12pt]{article}
\newfont{\bbd}{msbm10 scaled\magstep1}

\begin{document}
\thispagestyle{empty}

\def\ve#1{\mid #1\rangle}
\def\vc#1{\langle #1\mid}

\newcommand{\p}[1]{(\ref{#1})}
\newcommand{\be}{\begin{equation}}
\newcommand{\ee}{\end{equation}}
\newcommand{\sect}[1]{\setcounter{equation}{0}\section{#1}}

\newcommand{\vs}[1]{\rule[- #1 mm]{0mm}{#1 mm}}
\newcommand{\hs}[1]{\hspace{#1mm}}
\newcommand{\mb}[1]{\hs{5}\mbox{#1}\hs{5}}
\newcommand{\Db}{{\overline D}}
\newcommand{\bea}{\begin{eqnarray}}

\newcommand{\eea}{\end{eqnarray}}
\newcommand{\wt}[1]{\widetilde{#1}}
\newcommand{\und}[1]{\underline{#1}}
\newcommand{\ov}[1]{\overline{#1}}
\newcommand{\sm}[2]{\frac{\mbox{\footnotesize #1}\vs{-2}}
           {\vs{-2}\mbox{\footnotesize #2}}}
\newcommand{\prt}{\partial}
\newcommand{\eps}{\epsilon}

\newcommand{\R}{\mbox{\rule{0.2mm}{2.8mm}\hspace{-1.5mm} R}}
\newcommand{\Z}{Z\hspace{-2mm}Z}

\newcommand{\cd}{{\cal D}}
\newcommand{\cg}{{\cal G}}
\newcommand{\ck}{{\cal K}}
\newcommand{\cw}{{\cal W}}

\newcommand{\vj}{\vec{J}}
\newcommand{\vl}{\vec{\lambda}}
\newcommand{\vz}{\vec{\sigma}}
\newcommand{\vt}{\vec{\tau}}
\newcommand{\vw}{\vec{W}}
\newcommand{\poiss}{\stackrel{\otimes}{,}}

\def\l#1#2{\raisebox{.2ex}{$\displaystyle
  \mathop{#1}^{{\scriptstyle #2}\rightarrow}$}}
\def\r#1#2{\raisebox{.2ex}{$\displaystyle
 \mathop{#1}^{\leftarrow {\scriptstyle #2}}$}}



\renewcommand{\thefootnote}{\fnsymbol{footnote}}
\newpage
\setcounter{page}{0}
\pagestyle{empty}
\begin{flushright}
{November 1999}\\
{LPENSL--TH--20/99}\\
{solv-int/9911005}
\end{flushright}
\vfill

\begin{center}
{\LARGE {\bf A note on real forms of the complex}}\\[0.3cm]
{\LARGE {\bf N=4 supersymmetric Toda chain hierarchy}}\\[0.3cm]
{\LARGE {\bf in real N=2 and N=4 superspaces}}\\[1cm]

{}~

{\large F. Delduc$^{a,1}$ and A. Sorin$^{b,2}$}
{}~\\
\quad \\
{\em {~$~^{(a)}$ Laboratoire de Physique$^\dagger$,
Groupe de Physique Th\'eorique,}}\\
{\em ENS Lyon, 46 All\'ee d'Italie, 69364 Lyon, France}\\[10pt]
{\em {~$~^{(b)}$ Bogoliubov Laboratory of Theoretical Physics, JINR,}}\\
{\em 141980 Dubna, Moscow Region, Russia}~\quad\\

\end{center}

\vfill

{}~

\centerline{{\bf Abstract}}
\noindent
Three inequivalent real forms of the complex N=4 supersymmetric
Toda chain hierarchy (Nucl. Phys. B558 (1999) 545, solv-int/9907004)
in the real $N=2$ superspace with one even and two odd real
coordinates are presented. It is demonstrated that the first of them
possesses a global $N=4$ supersymmetry, while the other two admit a twisted
$N=4$ supersymmetry. A new superfield basis in which supersymmetry
transformations are local is discussed and a manifest $N=4$ supersymmetric
representation of the $N=4$ Toda chain in terms of a chiral and an
anti-chiral $N=4$ superfield is constructed. Its relation to the complex 
$N=4$ supersymmetric KdV hierarchy is established. Darboux-Backlund
symmetries and a new real form of this last hierarchy possessing a
twisted $N=4$ supersymmetry are derived.

{}~

{}~



{}~

{}~

\vfill
{\em \noindent
1) E-Mail: francois.delduc@ens-lyon.fr\\
2) E-Mail: sorin@thsun1.jinr.ru }\\
$\dagger$) UMR 5672 du CNRS, associ\'ee \`a l'Ecole Normale Sup\'erieure de 
Lyon.
\newpage
\pagestyle{plain}
\renewcommand{\thefootnote}{\arabic{footnote}}
\setcounter{footnote}{0}

\noindent{\bf 1. Introduction.}
Recently the Lax pair representation of the even and odd
flows of the complex $N=4$ supersymmetric Toda chain hierarchy
in $N=2$ superspace were constructed in \cite{dgs}. The corresponding 
local and nonlocal Hamiltonians, the finite and infinite
discrete symmetries, the first two Hamiltonian structures and the
recursion operator connecting all evolution equations and the Hamiltonian
structures were also studied. The goal
of the present letter is first to analyse the possible real forms
of the $N=4$ Toda chain hierarchy in $N=2$ superspace, second to derive a 
manifest $N=4$ supersymmetric representation of its first nontrivial even 
flows in the real $N=4$ superspace, and third to clarify its relation
(if any) with the $N=4$ supersymmetric KdV hierarchy.

Let us start with a short summary of the results that we shall need
 concerning the complex $N=4$ supersymmetric
Toda chain hierarchy (see \cite{dgs,ls,ols,ols1} for more details).

The complex $N=4$ supersymmetric Toda chain hierarchy in the complex
$N=2$ superspace comprises an infinite set of even and odd
flows for two complex even $N=2$ superfields
$u(z,\theta^{+},\theta^{-})$ and $v(z,\theta^{+},\theta^{-})$, where
$z$ and $\theta^{\pm}$ are complex even and odd
coordinates, respectively. The flows are generated by complex
even and odd evolution derivatives
$\{{\textstyle{\partial\over\partial t_k}}, ~U^{\pm}_k, ~U_k,
~{\overline U}_k\}$ and $\{ D^{\pm}_k, ~Q^{\pm}_k\}$
($k \in \hbox{\bbd N}$), respectively, with the following length
dimensions:
\begin{eqnarray}
[{\textstyle{\partial\over\partial t_k}}]=[U^{\pm}_k]=
[U_k]=[{\overline U}_k]=-k, \quad
[D^{\pm}_k]=[Q^{\pm}_k]=-k+\frac{1}{2}.
\label{dimtimes}
\end{eqnarray}
The first few of these flows are :
\begin{eqnarray}
{\textstyle{\partial\over\partial t_1}}
\left(\begin{array}{cc} v\\ u \end{array}\right) =
{\partial}\left(\begin{array}{cc} v\\ u \end{array}\right),
\label{eqs1}
\end{eqnarray}
\begin{eqnarray}
&&{\textstyle{\partial\over\partial t_2}} v =
+v~'' +  2uv(D_{+}D_{-}v)-(D_{+}D_{-}v^2u)-v^2(D_{+}D_{-}u) +2v(uv)^2,
\nonumber\\
&&{\textstyle{\partial\over\partial t_2}} u =
-u~'' +  2uv(D_{+}D_{-}u)-(D_{+}D_{-}u^2v)-u^2(D_{+}D_{-}v) -2u(uv)^2,
\label{eqs}
\end{eqnarray}
\begin{eqnarray}
{\textstyle{\partial\over\partial t_3}} v &=&
v~''' -3(D_{+}v)~'(D_{-}uv)+3(D_{-}v)~'(D_{+}uv)
-3v~'(D_{+}u)(D_{-}v) \nonumber\\
&&+3v~'(D_{-}u)(D_{+}v)
-6vv~'(D_{+}D_{-}u)+6(uv)^2v~', \nonumber\\
{\textstyle{\partial\over\partial t_3}} u &=&
u~''' -3(D_{-}u)~'(D_{+}uv)+3(D_{+}u)~'(D_{-}uv)
-3u~'(D_{-}v)(D_{+}u) \nonumber\\ &&+3u~'(D_{+}v)(D_{-}u)
-6uu~'(D_{-}D_{+}v)+6(uv)^2u~',
\label{flow3}
\end{eqnarray}
\begin{eqnarray}
D^{\pm}_1 v= -D_{\pm}v\pm 2vD^{-1}_{\mp}(uv),
\quad D^{\pm}_1 u= -D_{\pm}u\mp 2uD^{-1}_{\mp}(uv),
\label{ff2-}
\end{eqnarray}
\begin{eqnarray}
\quad \quad Q^{\pm}_1 \left(\begin{array}{cc} v\\ u \end{array}\right) =
Q_{\pm}\left(\begin{array}{cc} v\\ u \end{array}\right),
\label{supersflows}
\end{eqnarray}
\begin{eqnarray}
\quad \quad U^{\pm}_0 v=\frac{1}{2}v-{\theta}^{\pm}
(D_{\pm}v\mp 2vD^{-1}_{\mp}(uv)), \quad U^{\pm}_0 u=\frac{1}{2}u -
{\theta}^{\pm} (D_{\pm}u\pm 2uD^{-1}_{\mp}(uv)),
\label{q1}
\end{eqnarray}
\begin{eqnarray}
&&U_0 v= -{\theta}^{+}(D_{-}v+ 2vD^{-1}_{+}(uv))
-{\theta}^{-}(D_{+}v- 2vD^{-1}_{-}(uv)), \nonumber\\
&&U_0 u=-{\theta}^{+}(D_{-}u- 2uD^{-1}_{+}(uv))
-{\theta}^{-}(D_{+}u+2uD^{-1}_{-}(uv)),
\label{qq1}
\end{eqnarray}
\begin{eqnarray}
\quad \quad \quad
{\overline U}_0 \left(\begin{array}{cc} v\\ u \end{array}\right) =
\frac{1}{2}~(~{\theta}^{-}(D_{+}+Q_{+})-{\theta}^{+}(D_{-}+Q_{-})~)
\left(\begin{array}{cc} v\\ u \end{array}\right).
\label{qqq1}
\end{eqnarray}
Throughout this letter, we shall use the notations $u'=\partial 
u={\partial\over\partial z}u$.
$D_{\pm}$ and $Q_{\pm}$ are odd covariant derivatives
and supersymmetry generators,
\begin{eqnarray}
D_{\pm}\equiv \frac{\partial}{\partial {\theta}^{\pm}} +
{\theta}^{\pm} {\partial}, \quad
Q_{\pm}\equiv \frac{\partial}{\partial {\theta}^{\pm}} -
{\theta}^{\pm} {\partial}.
\label{algDQ0}
\end{eqnarray}
They form the  algebra\footnote{Hereafter, we explicitly
present only
non-zero brackets.}
\begin{eqnarray}
\{ D_{\pm},D_{\pm}\} = +2{\partial}, \quad
\{ Q_{\pm},Q_{\pm}\} = -2{\partial}.
\label{alg00}
\end{eqnarray}
Using the explicit expressions of the flows (\ref{eqs1}--\ref{qqq1}), one
can calculate their algebra which has the following nonzero brackets:
\begin{eqnarray}
\Bigl\{D^{\pm}_1\,,\,D^{\pm}_1\Bigr\}=
-2\;\frac{{\partial}}{{\partial t_{1}}}, \quad
\Bigl\{Q^{\pm}_1\,,\,Q^{\pm}_1\Bigr\}=
+2\;\frac{{\partial}}{{\partial t_{1}}},
\label{alg1}
\end{eqnarray}
\begin{eqnarray}
\Bigl[U_0\,,\,U^{\pm}_0\Bigr]= \pm {\overline U}_{0}, \quad
\Bigl[ {\overline U}_0\,,\,U^{\pm}_0\Bigr]=\pm U_{0}, \quad
\Bigl[U_0\,,\,{\overline U}_0\Bigr]=2 (U^{+}_{0}-U^{-}_{0}),
\label{algq}
\end{eqnarray}
\begin{eqnarray}
&&\Bigl[U^{\pm}_0\,,\,D^{\pm}_1\Bigr]=-Q^{\pm}_{1}, \quad
\Bigl[U^{\pm}_0\,,\,Q^{\pm}_1\Bigr]=-D^{\pm}_{1}, \nonumber\\
&&\Bigl[U_0\,,\,D^{\pm}_1\Bigr]=+Q^{\mp}_{1}, \quad
\Bigl[U_0\,,\,Q^{\pm}_1\Bigr]=+D^{\mp}_{1}, \nonumber\\
&&\Bigl[{\overline U}_0\,,\,D^{\pm}_1\Bigr]=\pm D^{\mp}_{1}, \quad
\Bigl[{\overline U}_0\,,\, Q^{\pm}_1\Bigr]=\pm Q^{\mp}_{1}.
\label{algqqbar}
\end{eqnarray}
This algebra reproduces the algebra of the global complex $N=4$
supersymmetry, together with its $gl(2,\hbox{\bbd C})$ automorphisms. 
It is the algebra of symmetries of the nonlinear even flows
(\ref{eqs}--\ref{flow3}). It may be realized in the superspace
$\{t_k,\theta^{\pm}_k,\rho^{\pm}_k, h^{\pm}_k,h_k, {\overline h}_k \}$,
where $t_k, h^{\pm}_k,h_k,{\overline h}_k$
($\theta^{\pm}_k,\rho^{\pm}_k$) are complex even (odd) abelian
evolution times with the length dimensions
\begin{eqnarray}
[t_k]=[h^{\pm}_k]=[h_k]=[{\overline h}_k]=k, \quad
[\theta^{\pm}_k] =[\rho^{\pm}_k]=k-\frac{1}{2}
\label{dim}
\end{eqnarray}
which are in one-to-one correspondence with the length dimensions
\p{dimtimes} of the corresponding evolution derivatives.

{}~

\noindent{\bf 2. Real forms of the N=4 Toda chain hierarchy.}
It is well known that different real forms derived from the same complex
integrable hierarchy are inequivalent in general.
Keeping this in mind it seems important to find as many
different real forms of the $N=4$ Toda chain hierarchy as possible.

With this aim let us discuss various inequivalent complex
conjugations of the superfields $u(z,\theta^{+},\theta^{-})$ and
$v(z,\theta^{+},\theta^{-})$, of the superspace coordinates 
$\{z,~\theta^{\pm}\}$,
and of the evolution derivatives
$\{{\textstyle{\partial\over\partial t_k}},
~U^{\pm}_k, ~U_k,~{\overline U}_k,~D^{\pm}_k, ~Q^{\pm}_k\}$
which should be consistent with the flows (\ref{eqs1}--\ref{qqq1}).
We restrict our considerations to the case
when $iz$ and $\theta^{\pm}$ are coordinates of the real $N=2$ superspace
which satisfy the following standard complex conjugation properties:
\begin{eqnarray}
(iz,{{\theta}^{\pm}})^{*}=(iz,{\theta}^{\pm}),
\label{conj}
\end{eqnarray}
where $i$ is the imaginary unity. We will also use the standard
convention regarding complex conjugation of products involving
odd
operators and functions (see, e.g., the books \cite{ggrs}).
In particular, if ${\cal O}$ is some even differential operator
acting on a superfield $F$, we define the complex conjugate of 
${\cal O}$ by $({\cal O}F)^*={\cal O}^*F^*$.  Then, in the
case under consideration one can derive, for example, the following
relations
\begin{eqnarray}
{\partial}^*=-{\partial}, \quad
({\epsilon}^{\pm})^{*}={\epsilon}^{\pm}, \quad
({\epsilon}^{+}{\epsilon}^{-})^{*}=-{\epsilon}^{+}{\epsilon}^{-},\quad
({\epsilon}^{\pm}D_{\pm})^{*}={\epsilon}^{\pm}D_{\pm}, \quad
(D_{+}D_{-})^{*} = - D_{+}D_{-}
\label{conjrel}
\end{eqnarray}
which we use in what follows. Here, ${\epsilon}^{\pm}$ are constant
odd real parameters.

Direct verification shows that the flows (\ref{eqs1}--\ref{qqq1})
admit the three inequivalent complex conjugations:
\begin{eqnarray}
&& \quad \quad \quad \quad \quad (v,u)^{*}= (-v,u), \quad
(z,{{\theta}^{\pm}})^{*}=(-z,{\theta}^{\pm}), \nonumber\\
&&(t_p,U^{\pm}_p,U_p,{\overline U}_p,{\epsilon}^{\pm}_p D^{\pm}_p,
{\varepsilon}^{\pm}_pQ^{\pm}_p)^{*}=(-1)^{p}(t_p,U^{\pm}_p,U_p,
{\overline U}_p,-{\epsilon}^{\pm}_pD^{\pm}_p,
-{\varepsilon}^{\pm}_pQ^{\pm}_p), ~ ~ ~ ~ ~
\label{conj1}
\end{eqnarray}
\begin{eqnarray}
&&\quad \quad \quad \quad (v,u)^{\bullet}= (u,v), \quad
(z,{{\theta}^{\pm}})^{\bullet}=(-z,{\theta}^{\pm}), \nonumber\\
&&(t_p,U^{\pm}_p,U_p,{\overline U}_p,{\epsilon}^{\pm}_p D^{\pm}_p,
{\varepsilon}^{\pm}_pQ^{\pm}_p)^{\bullet}=(-t_p,U^{\pm}_p,U_p,{\overline U}_p,
{\epsilon}^{\pm}_p D^{\pm}_p,{\varepsilon}^{\pm}_pQ^{\pm}_p), ~ ~ ~ ~ ~
\label{conj2}
\end{eqnarray}
\begin{eqnarray}
&& \quad \quad (v,~u)^{\star}=(~-u(D_-D_+\ln u+uv),~ \frac{1}{u}~),
\quad (z,{{\theta}^{\pm}})^{\star}=(-z,{\theta}^{\pm}),\nonumber\\
&&(t_p,U^{\pm}_p,U_p,{\overline U}_p,{\epsilon}^{\pm}_p D^{\pm}_p,
{\varepsilon}^{\pm}_pQ^{\pm}_p)^{\star}=(-t_p,-U^{\pm}_p,
-U_p,{\overline U}_p,-{\epsilon}^{\pm}_p D^{\pm}_p,
{\varepsilon}^{\pm}_pQ^{\pm}_p),
\label{conj3}
\end{eqnarray}
where ${\epsilon}^{\pm}_p$ and ${\varepsilon}^{\pm}_p$
are constant odd real parameters. We would like to underline that
the complex conjugations of the evolution derivatives (the second
lines of eqs. (\ref{conj1}--\ref{conj3}) ) are defined and fixed
completely by the explicit expressions (\ref{eqs1}--\ref{qqq1})
for the flows.
These complex conjugations extract different real forms of the algebra
(\ref{alg1}--\ref{algqqbar}). The real forms of the algebra
(\ref{alg1}--\ref{algqqbar}) with the involutions
(\ref{conj1}--\ref{conj2}) correspond to a twisted real $N=4$
supersymmetry, while the real form corresponding to the involution
\p{conj3} reproduces the algebra of the real $N=4$ supersymmetry.
This last fact becomes more obvious if one uses the $N=2$ basis of 
the algebra with the generators
\begin{eqnarray}
&& \quad \quad \quad
\Sigma_1\equiv U_0, \quad \Sigma_2 \equiv -i{\overline U}_0, \quad
\Sigma_3\equiv U^{-}_0-U^{+}_0, \quad \Sigma \equiv U^{-}_0+U^{+}_0,
\nonumber\\ &&{\cal D}_1\equiv Q^{+}_1+D^{+}_1, \quad
{\cal D}_2\equiv Q^{-}_1+D^{-}_1, \quad
{\overline {\cal D}}^{1}\equiv Q^{+}_1-D^{+}_1, \quad
{\overline {\cal D}}^{2}\equiv Q^{-}_1-D^{-}_1.
\label{su(2)}
\end{eqnarray}
Then, the nonzero algebra brackets (\ref{alg1}--\ref{algqqbar}) and the
complex conjugation rule \p{conj3} are the 
standard ones for the real  $N=4$ supersymmetry
algebra together with its $u(2)$ automorphisms
\begin{eqnarray}
&&\Bigl\{{\cal D}_{\alpha}\,,\,{\overline {\cal D}}^{\beta}\Bigr\}=
4 {{\delta}_{\alpha}}^{\beta}
\;\frac{{\partial}}{{\partial t_{1}}},\quad
\Bigl[{\Sigma}_a\,,\,{\Sigma}_b\Bigr]= 2i {\epsilon}_{abc}\Sigma_c,
\nonumber\\
&&\Bigl[{\Sigma}_a\,,\,{\cal D}_{\alpha}\Bigr]=
({{\sigma}_a)_{{\alpha}}}^{{\beta}}{\cal D}_{\beta}, \quad
~\Bigl[{\Sigma}_a\,,\,{\overline {\cal D}}^{\alpha}\Bigr]=
-{\overline {\cal D}}^{\beta}({{\sigma}_a)_{{\beta}}}^{{\alpha}},
\nonumber\\
&&\Bigl[{\Sigma}\,,\,{\cal D}_{\alpha}\Bigr]=
-{\cal D}_{\alpha}, \quad \quad \quad
~~\Bigl[{\Sigma}\,,\,{\overline {\cal D}}_{\alpha}\Bigr]=
{\overline {\cal D}}_{\alpha},
\label{algnn2}
\end{eqnarray}
\begin{eqnarray}
({\textstyle{\partial\over\partial t_1}},
{\Sigma}_a,{\Sigma},{\epsilon}^{\alpha}{\cal D}_{\alpha},
{\overline {\epsilon}}_{\alpha}
{\overline {\cal D}}^{\alpha})^{\star}=
-({\textstyle{\partial\over\partial t_1}},
 {\Sigma}_a,{\Sigma},-{\overline {\epsilon}}_{\alpha}
{\overline {\cal D}}^{\alpha},
-{\epsilon}^{\alpha}{\cal D}_{\alpha}), \quad
({\epsilon}^{\alpha},{\overline {\epsilon}}_{\alpha})^{\star}=
({\overline {\epsilon}}_{\alpha},{\epsilon}^{\alpha}).
\label{conjn23}
\end{eqnarray}
Here, ${\epsilon}^{\alpha},
{\overline {\epsilon}}_{\alpha}$ are constant odd parameters,
a summation over repeated indices
${\alpha}, {\beta}=1,2$ and $a,b,c=1,2,3$ is understood in eqs.
\p{algnn2}, and ${\sigma}_a$ are the Pauli matrices
\begin{eqnarray}
\sigma_1\equiv
\left(\begin{array}{cc} 0 & 1\\ 1&0 \end{array}\right), \quad
\sigma_2 \equiv \left(\begin{array}{cc} 0 & -i\\ i&0 \end{array}\right),
\quad \sigma_3\equiv \left(\begin{array}{cc} 1 & 0\\ 0&-1
\end{array}\right), \quad
{\sigma}_a {\sigma}_b= {\delta}_{ab}+
i {\epsilon}_{abc}\sigma_c,
\label{sigma}
\end{eqnarray}
and ${\epsilon}_{abc}$ is a totally antisymmetric tensor
(${\epsilon}_{123}=1)$. Therefore, we conclude that the complex $N=4$
supersymmetric Toda chain hierarchy with the complex conjugation \p{conj3}
possesses a real $N=4$ supersymmetry, and due to this
remarkable fact it can be called the $N=4$ supersymmetric Toda chain
hierarchy (for some examples of $N=4$ supersymmetric integrable
systems, see \cite{di}--\cite{zp} and references therein).

Let us remark that a combination of the two involutions
(\ref{conj3}) and (\ref{conj2}) generates the infinite-dimensional group
of discrete Darboux transformations \cite{ls,ols,dgs}
\begin{eqnarray}
&&(v,~u)^{\star \bullet}=(~v(D_-D_+\ln v-uv),~ \frac{1}{v}~), \quad
(z,{{\theta}^{\pm}})^{\star \bullet}=(z,{\theta}^{\pm}),\nonumber\\
&&(t_p,U^{\pm}_p,U_p,{\overline U}_p,D^{\pm}_p,
Q^{\pm}_p)^{\star \bullet }=(t_p,-U^{\pm}_p,
-U_p,{\overline U}_p,-D^{\pm}_p,Q^{\pm}_p).
\label{discrsymm}
\end{eqnarray}
This way of deriving discrete symmetries was proposed
in \cite{s} and applied to the construction of discrete symmetry
transformations of the $N=2$ supersymmetric GNLS hierarchies.

{}~

\noindent{\bf 3. A KdV-like basis with locally realized
supersymmetries.}
The third complex conjugation \p{conj3} looks rather complicated
when compared to the first two ones (\ref{conj1}--\ref{conj2}).
However, in another superfield basis defined as
\begin{eqnarray}
J\equiv uv + D_-D_+\ln u, \quad {\overline J}\equiv -uv,
\label{basis}
\end{eqnarray}
where $J$ and ${\overline J}$ ($[J]=[{\overline J}]=-1$) are 
unconstrained even $N=2$ superfields, it drastically simplifies.
In this basis the complex conjugations (\ref{conj1}--\ref{conj3}) are
given by
\begin{eqnarray}
(J,~{\overline J})^{*}=-(J,~{\overline J}),
\label{conj1j}
\end{eqnarray}
\begin{eqnarray}
(J,~{\overline J})^{\bullet}=(~J- D_-D_+\ln {\overline J},~{\overline J}~),
\label{conj2j}
\end{eqnarray}
\begin{eqnarray}
(J,~{\overline J})^{\star}=({\overline J},~J),
\label{conj3j}
\end{eqnarray}
and the equations (\ref{eqs}--\ref{qqq1})
become simpler as well,
\begin{eqnarray}
&&{\textstyle{\partial\over\partial t_2}} J =
-J~'' -2(J D_{+}^{-1}D_{-}{\overline J})~'+
D_{+}D_{-}(D_{+}^{-1}D_{-}J)^2, \nonumber\\
&&{\textstyle{\partial\over\partial t_2}}{\overline J} =
+{\overline J}~''  -2({\overline J}D_{+}^{-1}D_{-}J)~'+
D_{+}D_{-}(D_{+}^{-1}D_{-}{\overline J})^2,
\label{eqs2j}
\end{eqnarray}
\begin{eqnarray}
{\textstyle{\partial\over\partial t_3}} J &=&
J~''' +3\Bigl[
J~'D_{+}^{-1}D_{-}(J+{\overline J})+
J(D_{+}^{-1}D_{-}(J+{\overline J}))^2
+(D_{-}J)D_{+}J-{\overline J}J^2-
\frac{1}{3}J^3\Bigr]~'\nonumber\\
&-&3D_{+}D_{-}
\Bigl[J^2D_{+}^{-1}D_{-}(J+{\overline J})\Bigr], \nonumber\\
{\textstyle{\partial\over\partial t_3}} {\overline J} &=&
{\overline J}~''' -3\Bigl[
{\overline J}~'D_{+}^{-1}D_{-}(J+{\overline J})-
{\overline J}(D_{+}^{-1}D_{-}(J+{\overline J}))^2
+(D_{-}{\overline J})D_{+}{\overline J}+J{\overline J}^2+
\frac{1}{3}{\overline J}^3\Bigr]~'\nonumber\\
&-&3D_{+}D_{-}
\Bigl[{\overline J}^2D_{+}^{-1}D_{-}(J+{\overline J})\Bigr],
\label{eqs2jt3}
\end{eqnarray}
\begin{eqnarray}
\quad \quad \quad D^{\pm}_1
\left(\begin{array}{cc} J\\ {\overline J} \end{array}\right) =
D_{\pm}\left(\begin{array}{cc} +J\\ -{\overline J} \end{array}\right),
\quad Q^{\pm}_1
\left(\begin{array}{cc} J \\ {\overline J}  \end{array}\right) =
Q_{\pm}\left(\begin{array}{cc} J \\ {\overline J}  \end{array}\right),
\label{supersflowsj}
\end{eqnarray}
\begin{eqnarray}
\quad \quad \quad
U^{\pm}_0 \left(\begin{array}{cc} J \\ {\overline J}  \end{array}\right) =
-D_{\pm}{\theta}^{\pm}
\left(\begin{array}{cc} +J \\ -{\overline J}  \end{array}\right),
\quad  U_0 \left(\begin{array}{cc} J \\ {\overline J}
\end{array}\right)
= ({\theta}^{+}D_{-}+{\theta}^{-}D_{+})
\left(\begin{array}{cc} +J \\ - {\overline J} \end{array}\right),
\label{qqj1}
\end{eqnarray}
\begin{eqnarray}
\quad \quad {\overline U}_0 \left(\begin{array}{cc} J \\ {\overline J}
\end{array}\right) =
\frac{1}{2}~(~{\theta}^{-}(D_{+}+Q_{+})-{\theta}^{+}(D_{-}+Q_{-})~)
\left(\begin{array}{cc} J\\ {\overline J} \end{array}\right).
\label{qqqj1}
\end{eqnarray}
Notice that the supersymmetry  and $u(2)$ 
transformations (\ref{supersflowsj}--\ref{qqqj1}) of the superfields
$J$, $\bar J$ are local functions of the superfields, 
while the evolution equations (\ref{eqs2j}--\ref{eqs2jt3}) become nonlocal.

{}~

\noindent{\bf 4. A manifest $N=4$ supersymmetric representation.}
The equations (\ref{eqs2j}--\ref{eqs2jt3}) admit a manifestly $N=4$
supersymmetric
representation
\begin{eqnarray}
&&{\textstyle{\partial\over\partial t_2}} {\cal J} =
-{\cal J}~'' -{\cal D}_{+}{\cal D}_{-}\Bigl[2({\cal J} {\partial}^{-1}
{\overline {\cal J} })~'- ({\overline {\cal D}}^{+}
{\overline {\cal D}}^{-}{\partial}^{-1}{\cal J} )^2\Bigr], \nonumber\\
&&{\textstyle{\partial\over\partial t_2}}{\overline {\cal J}} =
+{\overline {\cal J}}~''  - {\overline {\cal D}}^{+}
{\overline {\cal D}}^{-}
\Bigl[2({\overline {\cal J}}{\partial}^{-1}{\cal J})~'-
({\cal D}_{+}{\cal D}_{-}{\partial}^{-1}{\overline {\cal J}})^2\Bigr],
\label{eqs2jN=4}
\end{eqnarray}
\begin{eqnarray}
{\textstyle{\partial\over\partial t_3}} {\cal J} &=&
{\cal J}~''' +{\cal D}_{+}{\cal D}_{-}\Bigl\{3\Bigl[
{\cal J}~'{\partial}^{-1}{\overline {\cal J}} +
({\cal J}{\partial}^{-1}{\overline {\cal J}})
{\cal D}_{+} {\cal D}_{-}{\partial}^{-1}{\overline {\cal J}}
-\frac{1}{2}({\overline {\cal D}}^{+}{\overline {\cal D}}^{-}
{\partial}^{-1}{\cal J})^2\Bigr]~'\nonumber\\
&-&({\overline {\cal D}}^{+}{\overline {\cal D}}^{-}{\partial}^{-1}
{\cal J})^3 -3({\overline {\cal D}}^{+}{\overline {\cal D}}^{-}
{\partial}^{-1}{\cal J})^2
{\cal D}_{+}{\cal D}_{-}{\partial}^{-1}{\overline {\cal J}}+
6{\cal J}{\overline {\cal J}}
~{\overline {\cal D}}^{+} {\overline {\cal D}}^{-}{\partial}^{-1}
{\cal J}\Bigr\}, \nonumber\\
{\textstyle{\partial\over\partial t_3}}{\overline {\cal J}} &=&
{\overline {\cal J}}~'''  +
{\overline {\cal D}}^{+}{\overline {\cal D}}^{-}\Bigl\{3\Bigl[
-{\overline {\cal J}}~'{\partial}^{-1}{\cal J} +
({\overline {\cal J}}{\partial}^{-1}{\cal J})
{\overline {\cal D}}^{+} {\overline {\cal D}}^{-}{\partial}^{-1}{\cal J}
+\frac{1}{2}({\cal D}_{+}{\cal D}_{-}
{\partial}^{-1}{\overline {\cal J}})^2\Bigr]~'\nonumber\\
&-&({\cal D}_{+}{\cal D}_{-}{\partial}^{-1}
{\overline {\cal J}})^3 -3
({\cal D}_{+}{\cal D}_{-}{\partial}^{-1}{\overline {\cal J}})^2
{\overline {\cal D}}^{+}{\overline {\cal D}}^{-} {\partial}^{-1}{\cal J}+
6{\cal J}{\overline {\cal J}}
~{\cal D}_{+}{\cal D}_{-}{\partial}^{-1}{\overline {\cal J}}\Bigr\},
\label{eqs2jN=4t3}
\end{eqnarray}
in terms of one chiral 
${\cal J}(z,\theta^+,\theta^-,\eta^+,\eta^-)$ and 
one antichiral
${\overline {\cal J}}(z,\theta^+,\theta^-,\eta^+,\eta^-)$
even $N=4$ superfield,
\begin{eqnarray}
{\cal D}_{\pm}{\cal J} =0, \quad
{\overline {\cal D}}^{\pm} ~{\overline {\cal J}} = 0.
\label{N=4constr}
\end{eqnarray}
Here ${\cal D}_{\pm},{\overline {\cal D}}_{\pm}$ are $N=4$ odd
covariant derivatives,
\begin{eqnarray}
&&{\cal D}_{\pm}\equiv \frac{1}{2}(
\frac{\partial}{\partial {\theta}^{\pm}} +
i\frac{\partial}{\partial {\eta}^{\pm}}+
({\theta}^{\pm} +i{\eta}^{\pm}) {\partial}), \quad
{\overline {\cal D}}^{\pm}\equiv \frac{1}{2}(
\frac{\partial}{\partial {\theta}^{\pm}} -
i\frac{\partial}{\partial {\eta}^{\pm}}+
({\theta}^{\pm} -i{\eta}^{\pm}) {\partial}), \nonumber\\
&&\quad \quad \quad \Bigl\{{\cal D}_{k}\,,\,
{\overline {\cal D}}^{m}\Bigr\}=
{{\delta}_{k}}^{m} {\partial}, \quad
\Bigl\{{\cal D}_{k}\,,\,{\cal D}_{m}\Bigr\}=
\Bigl\{{\overline {\cal D}}^{k}\,,\,
{\overline {\cal D}}^{m}\Bigr\}=0,\quad k,m=\pm,
\label{algnn4}
\end{eqnarray}
and $\eta^{\pm}$ are two additional real odd coordinates. The
relations between the independent components of the $N=2$ superfields
$\{J(z,\theta^+,\theta^-),~ {\overline J}(z,\theta^+,\theta^-)\}$
and those of the $N=4$ superfields
${\{\cal J}(z,\theta^+,\theta^-,\eta^+,\eta^-),
~{\overline {\cal J}}(z,\theta^+,\theta^-,\eta^+,\eta^-)\}$ are the
following:
\begin{eqnarray}
&&{\cal J}|_{\eta_{\pm} =0} =J, \quad
{\overline {\cal D}}^{\pm} {\cal J}|_{\eta_{\pm} =0} = D_{\pm}J, \quad
{\overline {\cal D}}^{+}~ {\overline {\cal D}}^{-}
{\cal J}|_{\eta_{\pm} =0} = D_+D_{-}J,
\nonumber\\ &&{\overline {\cal J}}|_{\eta_{\pm} =0} ={\overline J}, \quad
{\cal D}_{\pm}{\overline {\cal J}}|_{\eta_{\pm} =0} =
D_{\pm}{\overline J}, \quad
{\cal D}_{+}~{\cal D}_{-} {\overline {\cal J}}|_{\eta_{\pm} =0} =
D_{+}D_{-}{\overline J}.
\label{N=4n2rel}
\end{eqnarray}

Let us also present a manifestly $N=4$
supersymmetric form of the discrete Darboux transformations
\begin{eqnarray}
{\cal J}^{\star \bullet \star \bullet}={\cal J} -
{\cal D}_-{\cal D}_+\ln {\overline {\cal J}}, \quad
{\overline {\cal J}}^{\star \bullet \star \bullet}={\overline {\cal J}} -
{\overline {\cal D}}^-{\overline {\cal D}}^+
\ln {\cal J}^{\star \bullet \star\bullet}
\label{conj2jn=4}
\end{eqnarray}
which can easily be derived using eqs. (\ref{conj2j}--\ref{conj3j}) and
\p{N=4n2rel}. They are discrete symmetries of the even and
odd flows of the $N=4$ supersymmetric Toda chain hierarchy.
In other words, if the set $\{{\cal J},{\overline {\cal J}} \}$ is
a solution of the $N=4$ Toda chain hierarchy, then the set
$\{{\cal J}^{\star \bullet \star \bullet},
{\overline {\cal J}}^{\star \bullet \star \bullet}\}$, related to
the former by eqs. \p{conj2jn=4}, is a solution of the hierarchy
as well. The equations \p{conj2jn=4} reproduce
the one-dimensional reduction of the two-dimensional $N=(2|2)$
superconformal Toda lattice \cite{eh,lds}.

Finally, we would like to remark that the equations
(\ref{eqs2jN=4}--\ref{eqs2jN=4t3})
can be rewritten in a local form, if one introduces a new
superfield basis defined by the following invertible transformations:
\begin{eqnarray}
&&{\cal J} \equiv {\cal D}_{+}{\overline \Psi}, \quad \quad
~{\overline {\cal J}}\equiv {\overline {\cal D}}^{-}\Psi,\nonumber\\
&&{\overline \Psi}\equiv {\overline {\cal D}}^{+}{\partial}^{-1}{\cal J},
\quad \Psi \equiv {\cal D}_{-} {\partial}^{-1}{\overline {\cal J}},
\label{basislocal}
\end{eqnarray}
where $ \Psi , {\overline \Psi}$ ($[\Psi]=[{\overline \Psi}]=-1/2$)
are new constrained odd $N=2$ superfields
\begin{eqnarray}
{\cal D}_{-}{\Psi} ={\overline {\cal D}}^{+}\Psi =0, \quad
{\cal D}_{-} {\overline {\Psi}} = 
{\overline {\cal D}}^{+} {\overline {\Psi}} = 0
\label{N=4constrpsi}
\end{eqnarray}
with the reality conditions which can be derived from eqs.
(\ref{conj1j}--\ref{conj3j}) and \p{basislocal}.
Then, these equations become
\begin{eqnarray}
&&{\textstyle{\partial\over\partial t_2}} \Psi =
+{\Psi}~'' +2{\cal D}_-{\overline {\cal D}}^-
({\overline \Psi}~{\overline {\cal D}}^{-}\Psi)-
{\overline {\cal D}}^{+}({\cal D}_+\Psi)^2, \nonumber\\
&&{\textstyle{\partial\over\partial t_2}} {\overline \Psi} =
-{\overline \Psi}~'' - 2{\overline {\cal D}}^+{\cal D}_+
({\Psi}~{\cal D}_{+}{\overline \Psi})+
{\cal D}_{-}({\overline {\cal D}}^-{\overline \Psi})^2,
\label{eqslocal}
\end{eqnarray}
\begin{eqnarray}
{\textstyle{\partial\over\partial t_3}} \Psi  &=&
{\Psi}~''' +3{\cal D}_-
\Bigl[({\overline {\cal D}}^-{\Psi})~'
~{\overline {\cal D}}^{-}{\overline \Psi}
+({\overline {\cal D}}^{-}{\Psi})({\overline {\cal D}}^{-}
{\overline \Psi})^2+
\frac{1}{2}{\overline {\cal D}}^{+}{\overline {\cal D}}^{-}
({\cal D}_{+}{\Psi})^{2}\Bigr] \nonumber\\
&+&{\overline {\cal D}}^{+}\Bigl[
({\cal D}_{+}{\Psi})^{3}-
3({\cal D}_{+}{\Psi})^{2}
{\overline {\cal D}}^{-}{\overline \Psi}-
6({\cal D}_{+}{\overline \Psi})
({\overline {\cal D}}^{-}{\Psi})
{\cal D}_{+}{\Psi}\Bigr], \nonumber\\
{\textstyle{\partial\over\partial t_3}} {\overline \Psi}  &=&
{\overline {\Psi}}~''' +3{\overline {\cal D}}^+
\Bigl[({\cal D}_+{\overline {\Psi}})~'~{\cal D}_{+}{\Psi}
+({\cal D}_{+}{\overline {\Psi}})({\cal D}_{+}{\Psi})^2-
\frac{1}{2}{\cal D}_{+}{\cal D}_{-}
({\overline {\cal D}}^{-}{\overline {\Psi}})^{2}\Bigr] \nonumber\\
&+&{\cal D}_{-}\Bigl[
({\overline {\cal D}}^{-}{\overline {\Psi}})^{3}-
3({\overline {\cal D}}^{-}{\overline \Psi})^{2}
{\cal D}_{+}{\Psi}-
6({\cal D}_{+}{\overline \Psi})
({\overline {\cal D}}^{-}{\Psi})
{\overline {\cal D}}^{-}{\overline {\Psi}}\Bigr].
\label{eqslocalt3}
\end{eqnarray}

{}~

\noindent{\bf 5. Relation with the $N=4$ supersymmetric KdV hierarchy.}
It is well known that there are often hidden relationships between
a priori unrelated hierarchies. Some examples are the $N=2$ NLS
and $N=2$ ${\alpha}=4$ KdV \cite{kst},  the "quasi" $N=4$ KdV and 
$N=2$ ${\alpha}= -2$ Boussinesq \cite{dgi}, the $N=2$ (1,1)-GNLS
and $N=4$ KdV \cite{s,bs}. These relationships may lead to a
deeper understanding of the hierarchies. They may help to
obtain a more complete description and to derive solutions. 
 
The existence of a manifestly $N=4$ supersymmetric, local form 
(\ref{eqslocal}--\ref{eqslocalt3}) of equations belonging to the $N=4$
supersymmetric Toda chain hierarchy, gives an additional evidence in favour
of the existence of a hidden relationship with the $N=4$ supersymmetric KdV
hierarchy \cite{di,dik}. 

It turns out that such a relationship indeed exists. Let us present it at
the level of the second flow equations \p{eqs2j} which in a new
superfield basis $\{ {\widetilde J},\Phi, {\overline \Phi}\}$ take the
following local form:
\begin{eqnarray}
&&-i{\textstyle{\partial\over\partial t_2}} {\widetilde J} =
-\frac{1}{2}(\Phi + {\overline \Phi})~'' -
2({\widetilde J} (\Phi - {\overline \Phi}))~'+
[D,{\overline D}]({\widetilde J} (\Phi + {\overline \Phi})), \nonumber\\
&&-i{\textstyle{\partial\over\partial t_2}}{\Phi} =
2D{\overline D}({\widetilde J}~' -{\widetilde J}^2
-\frac{3}{4} {\Phi}^2 +\frac{1}{2}{\Phi} {\overline \Phi}), \nonumber\\
&&-i{\textstyle{\partial\over\partial t_2}}{\overline \Phi} =
2{\overline D}D({\widetilde J}~' +{\widetilde J}^2
+\frac{3}{4} {\overline \Phi}^2 -\frac{1}{2}{\Phi} {\overline \Phi}),
\label{releq}
\end{eqnarray}
where ${\widetilde J},\Phi, {\overline \Phi}$ ($[{\widetilde J}]=[\Phi]=
[{\overline \Phi}]=-1$) are new unconstrained, chiral ($D ~{\Phi}=0$) and
antichiral (${\overline D}~ {\overline {\Phi}} =0$) even $N=2$ superfields,
respectively, related to the superfields $J, \overline J$ \p{basis} by
the following invertible transformations:
\begin{eqnarray}
&& \quad \quad \quad \quad
J\equiv \frac{1}{2}({\Phi}+{\overline {\Phi}})-i{\widetilde J},\quad
{\overline J}\equiv \frac{1}{2}({\Phi}+{\overline {\Phi}})+i{\widetilde J},
\nonumber\\ &&{\widetilde J} \equiv \frac{i}{2}(J-{\overline J}), \quad
\Phi \equiv D{\overline D}{\partial}^{-1}(J+{\overline J}), \quad 
{\overline \Phi} \equiv {\overline D}D{\partial}^{-1}(J+{\overline J}), 
\label{relch}
\end{eqnarray}
and $D, \overline D$ are $N=2$ odd covariant derivatives,
\begin{eqnarray}
&&D\equiv \frac{1}{2}(D_{+}+iD_{-}), \quad
{\overline D}\equiv \frac{1}{2}(D_{+}-iD_{-}), \nonumber\\
&& \Bigl\{D\,,\,{\overline D}\Bigr\}={\partial}, \quad
\Bigl\{D\,,\,D\Bigr\}=
\Bigl\{{\overline D}\,,\, {\overline D}\Bigr\}=0.
\label{relcovder}
\end{eqnarray}
Now, one can easily recognize that eq. \p{releq} is the second flow of 
the $N=4$ KdV hierarchy in a particular "$SU(2)$ frame"
(compare eqs. \p{releq} with eqs. (4.5) and (4.3c) from ref. \cite{dik}).
Moreover, in this basis the second Hamiltonian structure of the $N=4$ Toda
chain hierarchy \cite{dgs} reproduces the $N=4$ $SU(2)$ superconformal
algebra and the flow \p{releq} is generated by the Hamiltonian $H^{t}_2$
\cite{dgs} 
\begin{eqnarray}
H^{t}_2=\int dz d\theta^{+}d\theta^{-} uv~' \equiv 
i\int dz d\theta^{+}d\theta^{-} {\overline J}
[D,{\overline D}]{\partial}^{-1}(J+{\overline J}) \equiv  
\int dz d\theta^{+}d\theta^{-} {\widetilde J}
({\overline \Phi}-\Phi).  
\label{h2}
\end{eqnarray}
The same relationship is certainly valid for any other flow of the 
$N=4$ Toda and $N=4$ KdV hierarchies both in the $N=2$ and $N=4$
superspaces. 

The relationship just established allows to apply the  formalism
of ref. \cite{dgs}, developed for the case of the $N=4$ Toda chain
hierarchy, for a more complete description of the $N=4$ KdV hierarchy. It
can be used to construct new bosonic and fermionic flows and
Hamiltonians, new finite and infinite discrete symmetries, the tau function,
etc.. Let us present as an example the three involutions 
\begin{eqnarray}
(~\Phi,~\overline \Phi,~{\widetilde J}~)^{*}=
(~-\overline \Phi,~-\Phi,~{\widetilde J}~),
\label{conj1jkdv}
\end{eqnarray}
\begin{eqnarray}
&&\Phi^{\bullet}= \overline \Phi - i {\overline D}D
\ln (\Phi + \overline \Phi + 2i{\widetilde J}), \quad
{\overline \Phi}^{\bullet}= \Phi + i D{\overline D}
\ln (\Phi + \overline \Phi + 2i{\widetilde J}), \nonumber\\
&& \quad \quad \quad \quad \quad
{\widetilde J}^{\bullet}= -{\widetilde J} + \frac{1}{2} [D, {\overline D}]
\ln (\Phi + \overline \Phi + 2i{\widetilde J}),
\label{conj2jkdv}
\end{eqnarray}
\begin{eqnarray}
(~\Phi,~\overline \Phi,~{\widetilde J}~)^{\star}=
(~\overline \Phi,~\Phi,~{\widetilde J}~)
\label{conj3jkdv}
\end{eqnarray}
and the Darboux-Backlund symmetries
\begin{eqnarray}
&&\Phi^{\star \bullet}= \Phi + i D {\overline D}
\ln (\Phi + \overline \Phi + 2i{\widetilde J}), \quad
{\overline \Phi}^{\star \bullet}= \overline \Phi - i {\overline D}D
\ln (\Phi + \overline \Phi + 2i{\widetilde J}), \nonumber\\
&& \quad \quad \quad \quad \quad
{\widetilde J}^{\star \bullet}= -{\widetilde J} + \frac{1}{2} 
[D, {\overline D}] \ln (\Phi + \overline \Phi + 2i{\widetilde J})
\label{dbkdv}
\end{eqnarray}
derived using the relationship \p{relch} and eqs.    
(\ref{conj1j}--\ref{conj3j}). 
The involution \p{conj2jkdv} looks rather complicated, however 
let us remember that in the original superfield basis $\{u,v\}$ it has a
very simple form \p{conj2}. The involutions \p{conj1jkdv} and 
\p{conj3jkdv} were discussed in \cite{ik}, to our knowledge 
but the involution 
\p{conj2jkdv} and the corresponding real form of the $N=4$ KdV hierarchy
as well as its Darboux-Backlund symmetries \p{dbkdv}  
are presented here for the first time.

{}~

\noindent{\bf 6. Conclusion.}
In this letter, we have described three distinct real forms of the 
$N=4$ Toda chain hierarchy introduced in \cite{dgs}. It has been
shown that the symmetry algebra of one of these real forms contains 
the usual (untwisted) real $N=4$ supersymmetry algebra. A set of $N=2$ 
superfields with simple conjugation properties in the untwisted case 
have been introduced. It has then been shown how to extend these
superfields to superfields in $N=4$ superspace, and write all flows
and conjugation rules directly in $N=4$ superspace. Finally, a
change of basis in $N=4$ superspace has allowed us to eliminate
all nonlocalities in the flows. As a byproduct, a relationship between
the complex $N=4$ Toda chain and $N=4$ KdV hierarchies has been established, 
which allows to derive Darboux-Backlund symmetries and a new real form of the
last hierarchy possessing a twisted $N=4$ supersymmetry.

It is obvious that there remain a lot of work to do in order to improve
our understanding of the hierarchy in $N=4$ superspace. A first step in this 
direction would be to derive a Lax formulation of the flows in terms
of $N=4$ operators.

{}~

\noindent{\bf Acknowledgments.}
A.S. would like to thank L. Freidel for useful discussions and the
Laboratoire de Physique Th\'eorique - de l'ENS Lyon for the hospitality
during the course of this work.
This work was partially supported by the PICS Project No. 593,
RFBR-CNRS Grant No. 98-02-22034, RFBR Grant No. 99-02-18417,
Nato Grant No. PST.CLG 974874 and INTAS Grant INTAS-96-0538.

\end{document}